\documentclass[3p,12pt, times]{elsarticle}
\usepackage{amsmath}
\usepackage{amssymb}
\usepackage{algorithm}
\usepackage[noend]{algpseudocode}
\usepackage{graphicx}
\usepackage{multicol}
\usepackage{multirow}
\usepackage{float}
\usepackage{array}
\usepackage{siunitx}
\usepackage{caption}
\usepackage{subcaption}
\usepackage{adjustbox}
\usepackage{fontenc}
\usepackage{pifont}

\begin{document}
\begin{frontmatter}

\title{Evaluating Roles of Central Users in Online Communication Network: A Case Study of \#PanamaLeaks}
\author[label]{Mohsin Adalat} 
\author[label]{Muaz A. Niazi$^*$}

\address[label]{Computer Science Department,\\ COMSATS Institute of Information Technology,\\Islamabad, Pakistan\\ $^*$Corresponding author}

%

\begin{abstract}
Social media has changed the ways of communication, where everyone is equipped with the power to express their opinions to others in online discussion platforms. Previously, a number of studies have been presented to identify opinion leaders in online discussion networks. Feng (``Are you connected? Evaluating information cascade in online discussion about the \#RaceTogether campaign'', Computers in Human Behavior, $2016$) identified five types of central users and their communication patterns in an online communication network of a limited time span. However, to trace the change in communication pattern, a long-term analysis is required. In this study, we critically analyzed framework presented by Feng based on five types of central users in online communication network and their communication pattern in a long-term manner. We take another case study presented by Udnor et al. (``Determining social media impact on the politics of developing countries using social network analystics'', Program, $2016$) to further understand the dynamics as well as to perform validation . Results  indicate that there may not exist all of these central users in an online communication network in a long-term manner. Furthermore, we discuss the changing positions of opinion leaders and their power to keep isolates interested  in an online discussion network. 
 
\end{abstract}

\begin{keyword}
Complex network \sep Social network analysis \sep opinion formation

\end{keyword}

\end{frontmatter}

\section{Introduction}

To exchange opinions and information, internet is being used all-around the world. With increased use of internet, social media also became popular and a central medium for communication. In recent times, social networking has become de facto medium for communication among the same social group or network \cite{gole2015frequent}. People participate in online discussion by using different platforms like Facebook, Twitter, YouTube and other social networking tools \cite{himelboim2013birds}.Twitter is one of the popular social networking tool (micro blogs) among others, where people broadcast brief update messages \cite{martinez2016understanding}. 

In scientific literature, a number of studies have been carried out using Twitter data. For example, Ch'ng in \cite{ch2015bottom} used Twitter data to find generation of online communities and Baek in \cite{baek2015longitudinal} argued roles and positions of participants in an online community. Geo located tweets are used by Blanford in \cite{blanford2015geo} to capture regional connections and cross border movement. Twitter data is even used for evaluation of inter organizational disaster coordination networks, created in result of disaster \cite{abbasi2016longitudinal}. Lee et al. in \cite{lee2017mapping} demonstrated how scholars are using Twitter communication network for informal scholarly communication. Authors used social network analysis methods in all of the above-mentioned studies. Social network analysis is a strategy for analyzing social structures \cite{otte2002social}. Aim of all the above studies in different fields is to find the influential nodes in network. In online communication network, these nodes are called opinion leaders. 

Rogers defined opinion leadership as ``the degree to which an individual is able to influence other individuals' attitudes or overt behavior informally in desired way with relative frequency'' \cite{rogers2003diffusion}. The concept of opinion leadership originates from two step flow theory, the theory posits that information first comes from mass media to opinion leaders and then this information is transmitted to society via opinion leaders \cite{katz1955lazarsfeld}. Opinion leaders are opinion broker in two sense, first opinion leaders influence is between two social groups rather than in-between social group and second they are transition between two network mechanism responsible for spreading the idea \cite{burt1999social}.

For identification of opinion leaders in Twitter network, Borge in \cite{borge2017opinion} used social network analysis techniques on following-follower and mention network. Some researchers proposed different algorithms to find opinion leaders in social networks. For example, Aleahmad et al. proposed OLFinder in  \cite{aleahmad2016olfinder} and Huang proposed a Dynamic Opinion Rank algorithm in \cite{huang2014finding}. Li et al. in \cite{li2011talking} proposed framework for opinion leader identification, which consist of five stages. First is keyword blog search; second is ontology extraction; ontology-assisted extraction and hot blog identification is third and fourth respectively; fifth and last stage of framework is opinion leader identification. The method of identifying opinion leaders in virtual communities proposed by Zhang in \cite{zhang2009ways} consist of five steps: first, find twenty participators who are active with large followers; second, to investigate details of these active users, construct a relational matrix; third, centrality analysis; fourth, identify opinion leaders by comparing standardized degree of network users; last, if centrality and map analysis fails then correlation analysis is used. Researchers also applied two step flow theory \cite{katz1955lazarsfeld} for identifying opinion leaders in online social network. 

By adopting two step flow theory, Feng in \cite{feng2016you} proposed another approach to identify central users in the online communication network. In this work, the author identified five types of central users in Twitter network: conversation starter, influencer, active engager, network builder, and information bridge. However, Feng study analyzed tweets in only four-days time span to trace communication patterns that is not appropriate because opinion in a society may vary with time. The proposed approach needs to be applied for trend analysis on longitudinal network.

In this context, this study explores communication patterns among participants of online discussion network in longitudinal manner by adopting previous approach  proposed by Feng in \cite{feng2016you}. In this study, we use  ``PanamaLeaks'' as  case study. International Consortium of Investigative Journalists' raised issue of Panama Leaks by leaking 11.1 million records from Panama-based law firm, Mossack Fonseca \cite{cross2017panama}. We retrieve Twitter data using \#PanamaLeaks keyword from April 2017 to August 2017. The research questions we address are: Weather central users identified by Feng study exist in online communication network in longitudinal manner?.  If yes, then what are the changes in communication pattern of central users?. For validation, we apply the same approach on another case study of \#NeigieraDecides presented in \cite{udanor2016determining}.

Our main contributions can be listed as follows:
\begin{itemize}
\item To identify most central users in longitudinal network of \#PanamaLeaks.
\item To analyze the flow of information among the participants of online longitudinal discussion network in \#PanamaLeaks.
\end{itemize}
The reminder of the paper is structured as follows: Section 2 presents background, Section 3 presents our proposed method and dataset followed by results presented in section 4, Discussion is formulated in section 5.

\section{Background}
In this section, we present characteristics and  communication patterns of the most central users identified by Feng in \cite{feng2016you}.
\subsection{Conversation starter}

\begin{figure}[H]
\captionsetup{justification = centering}
\begin{center}
\includegraphics[width=5.0 cm, height=5.0 cm]{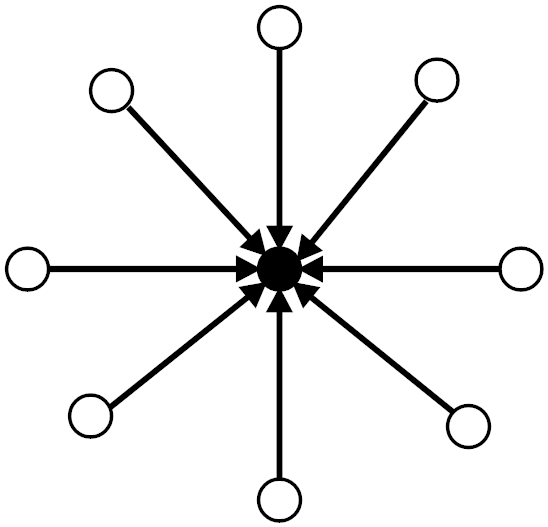}
\caption{Connection Pattern for ``Conversation Starter'' in online discussion network.Circles are presenting users in the network. Circle filled with black color is ``Conversation Starter''. Many arrows pointing towards ``Conversation Starter'' presents many users in the network mentioning/retweeting ``Conversation Starter'' tweet. }\label{fig:cs}
\end{center}
\end{figure}

A \emph{conversation starter} is a user in network with numerus ``in-degree'' links and a few or none ``out-degree'' links. \textit{Conversation starter} is the one, who is responsible for starting the original topic and flow of information in the network. However, the control on flow of information in network is not under control of the conversation starter. Figure \ref{fig:cs} shows connection pattern for \textit{conversation starter} in the network. 
\subsection{Influencer}
\begin{figure}[H]
\captionsetup{justification = centering}
\begin{center}
\includegraphics[width=5.0 cm, height=5.0 cm]{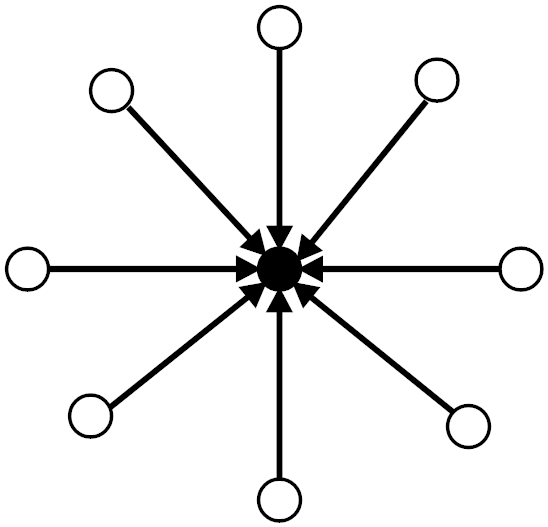}
\caption{Connection Pattern for ``Influencer'' in online discussion network. Circles are presenting users in the network. Circle filled with black color is ``influencer''. Many arrows pointing towards ``Influencer'' presents many users in the network mentioning/retweeting ``Influencer'' tweet. }\label{fig:inf}
\end{center}
\end{figure}

An \textit{influencer} is opinion leader in the network. \textit{Influencer} in a network has plentiful ``in-degree'' links and few ``out-degree'' links.\textit{ Influencer} does not initiate the topic of conversation in the network, however \textit{influencer} do influence opinion of other users in the network by creating frequent tweets that are retweeted by other users. Figure \ref{fig:inf} shows connection pattern for \textit{influencer} in the network. 

\subsection{Active engager}
\begin{figure}[H]
\captionsetup{justification = centering}
\begin{center}
\includegraphics[width=5.0 cm, height=5.0 cm]{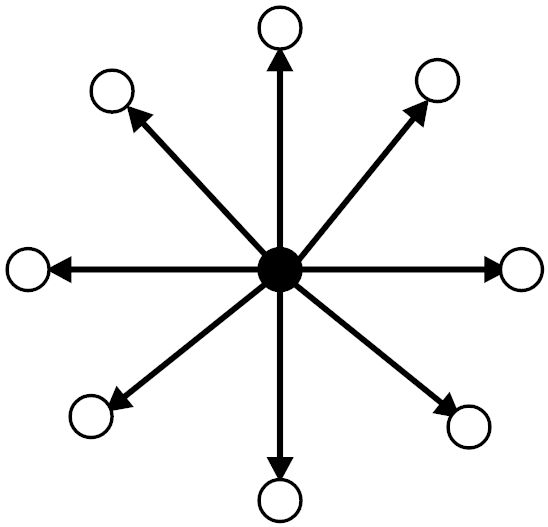}
\caption{Presenting Connection Pattern for ``Active Engager'' in online discussion network. Circles are presenting users in the network. Circle filled with black color is ``Active Engager''. Many arrows pointing towards users from ``Active Engager'' presents that ``Active Engager'' mention/retweet many other users tweet in the network. }\label{fig:ae}
\end{center}
\end{figure}

An \textit{active engager} in a user in online discussion network with many ``out-degree'' and a few or none ``in-degree'' links. \textit{Active engager} is eager to distribute information and make connections in online discussion network. \textit{Active engager} is also opinion expresser in network. Figure \ref{fig:ae} shows connection pattern for \textit{active engager} in the network. 

\subsection{Network builder}
\begin{figure}[H]
\captionsetup{justification = centering}
\begin{center}
\includegraphics[width=9.0 cm, height=5 cm]{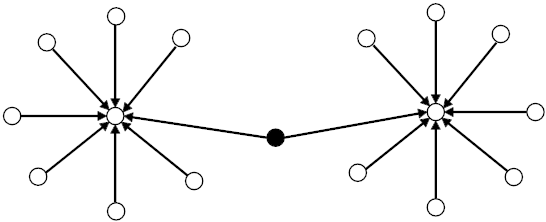}
\caption{Connection Pattern for ``Network Builder'' in online discussion network. Circles are presenting users in the network. Circle filled with black color is ``network builder''. Two arrows pointing towards ``Influencers'' in the network is describing basic role of ``Network Builder'' in the network. }\label{fig:nb}
\end{center}
\end{figure}

Despite having a few ``out-degree'' and a few or none ``in-degree'' links in online discussion network, \textit{network builder} plays an important role in the network. Main role of \textit{network builder} is connecting two or more influencers in the network. Figure \ref{fig:nb} shows connection pattern for \textit{network builder} in the network.   

\subsection{Information bridge }
\begin{figure}[H]
\captionsetup{justification = centering}
\begin{center}
\includegraphics[width=9.0 cm, height=5.0 cm]{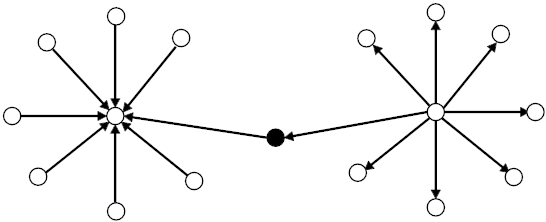}
\caption{Connection Pattern for ``Information Bridge'' in online discussion network. Circles are presenting users in the network. Circle filled with black color is ``Information Bridge''. Arrow pointing towards ``Influencer'' in the network is presenting that ``Information Bridge'' mention/retweet ``Influencer'' tweet and arrow from ``Active Engager'' towards ``Information Bridge'' is presenting that same tweet is retweeted by the ``Active Engager''.}\label{fig:ib}
\end{center}
\end{figure}

An \textit{information bridge} is a user in online discussion network with a few ``in-degree'' and ``out-degree'' links. \textit{Information bridge} role is to assist \textit{influencer} and \textit{active engager} in the network to connect with other users (Feng 2016). \textbf{Note} that in both case studies, \#Panamaleaks and \#NigeriaDecies which we took for validation. We are unable to find such user in the online discussion network (see section \ref{sec:disc}). Figure \ref{fig:ib} shows connection pattern for \textit{information bridge} in the network. 
\section{Proposed Methodology }

\subsection{Tool}
We choose NodeXl and Gephi as tools to analyze \#PanamaLeaks network. NodeXl is an extensible toolkit to download and analyze network data \cite{smith2009analyzing}. NodeXl is capable of calculating network matrices like ``in-degree'', ``out-degree'', ``betweeness'', ``density'', ``modularity'' etc. NodeXl generates excel-based reports and it is capable of downloading and analyzing data of several social network platforms like Twitter, Flickr and Facebook. In Twitter user network, a Twitter user is vertex in NodeXl and edge represents the relationship between users. Relationship between two vertices can be direct or un-directed. In Twitter user network, ``in-degree'' of each vertex is how many users in network have mentioned or retweeted his/her tweet. Retweets by a user in the network and other users mentioned by that specific user is out-degree of that vertex.

Along with finding the most central users of \#PanamaLeaks network, we aim to visualize \#PanamaLeaks network dynamically to find structural changes in the network. We choose Gephi social network analysis tool \cite{bastian2009gephi} for network visualization due to its capability for visualizing large, complex and dynamic networks. Gephi is capable of visualizing network with many different algorithm like Force-Atlas, Fruchterman-Reingold, Yifan Hu etc \cite{cherven2015mastering}.

\subsection{Data Source}
We used NodeXl Twitter Search Network data collector to get tweets having \#PanamaLeaks. It took us six months to collect these tweets because twitter streaming API only allows downloading previous seven days of tweets.We choose \#PanamaLeaks as keyword to download tweets. Our dataset contains 10612 nodes and 24623 edges exist in the \#Panamaleaks network during March 2017 to August 2017. 

For validation purpose, we took \#NigeriaDecides case study data set is used in \cite{udanor2016determining}. The author used \#NigeriaDecides as a query to download tweets and generated network of tweets, mentions and replies, which is same as \#PanamaLeaks case study. \#NigeriaDecides case study data set consist of 1752 nodes and 6343 edges from 15 April 2015 to 20 April 2015.

In both case studies  networks, nodes are the twitter users and edges are the relation between two twitter users. NodeXl add an edge, when one user tweets, retweets or mention other user/users in his/her tweet. Edges coming towards one users is ``in-degree'' of that specific user and edges going from that user to others in network are ``out-degree'' of that specific user. If a user creates new tweet/tweets without mentioning, replying or retweeting another user’s tweet, NodeXl adds self-loop around that user in the network. 

\subsection{Methods}
We combine all the data (Tweets downloaded using NodeXl during six months) and then divide it into six months of different networks according to the dates of tweets from March 2017 to August 2017. We requested NodeXl to generate following matrices ``in-degree'', ``out-degree'', ``betweenness centrality'', ``density'', ``clustering coefficient'', and ``modularity'' for each month independently. ``In-degree'' is simple count of unique incoming connections/links towards a user/entity in a directed graph. Whereas ``out-degree'' is simple count of unique links/connections from a user. ``Betweenness centrality'' of the node (A unique user in a network) is how many times a node appears in the shortest path between other nodes (other users in the same network) \cite{freeman1977set}. 

``In-degree'', ``out-degree'' and ``betweenness centrality'' are the three measures used by Feng in \cite{feng2016you} for identifying most central users in online discussion network. That is the reason we also used these three measures on \#PanamaLeaks case study. We identified top ten central users based on ``betweenness centrality'' for each months in the \#PanamaLeaks network. We also identified top ten central users in the \#NigeriaDecides case study used in \cite{udanor2016determining} for validation purpose. For visualization of connection patterns for most central users in \#PanamaLeaks network, we use Yifan Hu algorithm for both case studies. Figure \ref{fig:allresultscs} shows connection pattern for conversation starter for \#PanamaLeaks and \#NigeriaDecies. Figure \ref{fig:topinf}, \ref{fig:activeengager} and \ref{fig:networkbuilders} shows connection patterns for  \textit{influencer}, \textit{active engager} and \textit{network builder} in the \#PanamaLeaks  network for each month from April 2017 to August 2017. Figure \ref{fig:infdc}, \ref{fig:aedc} and \ref{fig:nbdc} shows the connection patterns for \#NigeriaDecides network.

We used social network analysis measures like ``density'', ``clustering coefficient'', and ``modularity'' to analyze overall communication pattern among the participants of the \#PanamaLeaks network. To check the impact of \textit{conversation starter} and \textit{influencer} on the online discussion network, we use number of unique users coming into \#PanamaLeaks network every month. 

\begin{figure}[H]
\begin{center}
\includegraphics[width=8.0 cm, height=8.0 cm]{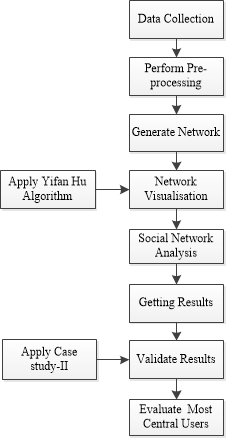}
\caption{Research method.}\label{fig:researchmethod}
\end{center}
\end{figure}

For visualization of overall communication pattern among \#PanamaLeaks network participant. We created dynamic network of \#PanamaLeaks using Gephi. We implemented month wise time stamp on the network based on tweet date.  We used Yifan Hu algorithm, one of the many available algorithm for network visualization in Gephi. Figure \ref{fig:strmarch}, \ref{fig:strapril}, \ref{fig:strmay}, \ref{fig:strjune}, \ref{fig:strjuly} and \ref{fig:straugust} are showing communication patterns between all users of \#PanamaLeaks network for March, April, May, June, July and August respectively. Figure \ref{fig:overalstructure} shows overall visualization of the complete six months network from March 2017 to August 2017 using Yifan Hu algorithm. Our overall research methodology has been shown in figure \ref{fig:researchmethod}.

\section{Results}
In this section, we present series of experiments carried out in our study.
\subsection{Conversation starter} 

According to the two-step flow theory, information flows from mass media to opinion leaders and then transmitted to the population via opinion leaders \cite{katz1955lazarsfeld}. In light of the above, we took ``siasatpk'' as our \textit{conversation starter} as it is a mass media platform. As per the result of NodeXl, \textit{conversation starter} topped the list as expected. According to results of March, April, May, June, and July ``in-degree'' of ``siasatpk'' is $ 211, 137, 203, 446$ and $1463 $ respectively. Whereas, ``Out-degree'' of ``siasatpk'' for March, April, May, June, and July is $1, 3, 2, 7$ and $22$ respectively. ``siasatpk'' who is playing the role as \textit{conversation starters} in \#PanamaLeaks network remains the first or second most influential participant of the network from March to July. ``In-degree'' of ``siasatpk'' is zero in the month of August because ``siasatpk'' stop taking participation in discussion from August (See Table \ref{tbl:rsltcs}).

\begin{figure}[H]
\captionsetup{justification=centering}
\begin{subfigure}{.5\textwidth}
  \centering
  \includegraphics[width=6.0 cm, height= 6.0 cm]{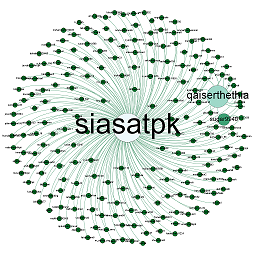}
  \caption{``Conversation Starter'' of March}
  \label{fig:csmarch}
\end{subfigure}%
\begin{subfigure}{.5\textwidth}
  \centering
  \includegraphics[width=6.0 cm, height= 6.0 cm]{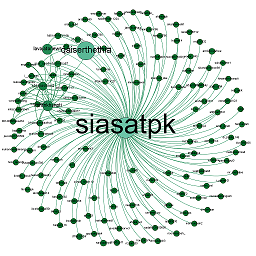}
  \caption{``Conversation Starter'' of April}
  \label{fig:csapril}
\end{subfigure}\hfill
\begin{subfigure}{.5\textwidth}
  \centering
  \includegraphics[width=6.0 cm, height= 6.0 cm]{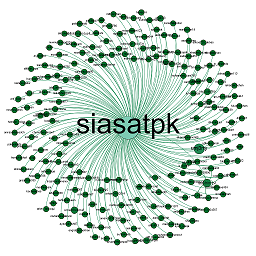}
  \caption{``Conversation Starter'' of May}
  \label{fig:csmay}
\end{subfigure}
\begin{subfigure}{.5\textwidth}
  \centering
  \includegraphics[width=6.0 cm, height= 6.0 cm]{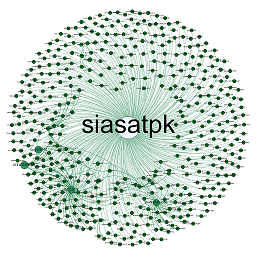}
  \caption{``Conversation Starter'' of June}
  \label{fig:csjune}
\end{subfigure} 
\begin{subfigure}{.5\textwidth}
\centering
  \includegraphics[width=6.0 cm, height= 6.0 cm]{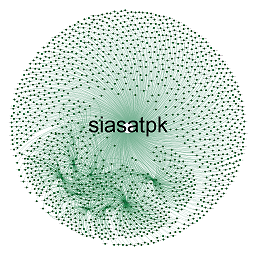}
  \caption{``Conversation Starter'' of July}
  \label{fig:csjuly}
\end{subfigure}
\begin{subfigure}{.5\textwidth}
  \centering
  \includegraphics[width=6.0 cm, height=6.0 cm]{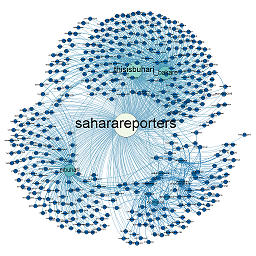}
  \caption{``Conversation Starter'' of the \#NigeriaDecides}
  \label{fig:csds}
\end{subfigure}

\caption{Connection Pattern for ``Conversation Starter'' of every month. Middle node of every figure is  ``Conversation Starter''. Size and color of nodes is according to their degree. Thus, the nodes bigger in size are high degree nodes and smaller nodes are nodes having low degree. Whereas, green nodes are nodes having lowest degree and color tends to change towards white with increase in degree. }
\label{fig:allresultscs}
\end{figure}

\begin{table}[H]
\captionsetup{justification = centering}
\caption{Presenting ``in-degree'', ``out-degree'', ``betweeness centrality'' and ``rank'' of ``siasatpk'' playing role as ``Conversation Starter'' in the \#Panamaleaks network. The number ``1'' in  rank column presents that ``conversation starter'' is the most central user in \#PanamaLeaks network.}\label{tbl:rsltcs}
\centering
\begin{tabular}{| p{2cm} |p{2cm } |p{3.0cm } |p{3.0cm }|p{1.5cm }|}
\hline

\hline
Month & In-degree & Out-degree & Betweenness & Rank \\
\hline

March &	211 &	1	& 492974.973 &	1 \\\hline
April &	137	& 3 &	461067.857 &	2 \\\hline
May	& 203	& 2 &	975715.667 &	2 \\\hline
June	& 446 &	7	& 1219538.696 & 1 \\\hline
July	& 1463	& 22 &	 7539507.557 &	1\\\hline
August	& 0& 	0&	-&	-\\\hline

\end{tabular}
\end{table}
 ``saharareporters'' with $( in-degree = 345, out-degree = 13$ and $betweenness-centrality = 616332.423)$ is playing the role of \textit{conversation starter} in the \#NigeriaDecides case study, which we used for validation. As far as the connection pattern for \textit{conversation starter} of \#PanamaLeaks network, Figures \ref{fig:csmarch},\ref{fig:csapril},\ref{fig:csmay},\ref{fig:csjune} and \ref{fig:csjuly} are showing the five months connection pattern of \textit{conversation starter}. Mostly isolates are pointing towards the \textit{conversation starter} (mentioning or retweeting tweets of conversation starter). \textit{Conversation starter} is acting as a hub in the \#PanamaLeaks network. Results of the \#NigeriaDecides case study used for validation revels the same pattern of communication between \textit{conversation starter} and other participants of the network. (See Figure \ref{fig:csds}).

\subsection{Influencer}
There can be more than one \textit{influencer} in the online discussion network as result shows. Here we present top three \textit{influencer} in \#PanamaLeaks network of each month based on ``betweenness centrality''. In March, ``qaiserthethia'' with $(in-degree = 139, out -degree = 12)$ is the third most central user of the network. Fourth most central user of the March network is ``alirazatweets'' with $(in-degree = 150, out -degree = 1)$. ``xaifoo\_official'' is the fifth most central user in March network with $(in-degree = 90, out -degree = 1)$.

In April, ``ptiofficial'' is the most central user of the network acting as an \textit{influencer} in the network with $(in-degree = 266, out -degree = 1)$ which means $266$ other users have mentioned or retweeted ``ptiofficial'' tweets. Third most central user is ``qaiserthethia'' with $(in-degree = 107, out -degree = 12)$ and ``amranafahad'' with $(in-degree = 116, out -degree = 2)$ is fourth most central user of April network. 

In May, the most central user is  ``khaleejmag'' with $(in-degree = 347, out -degree = 1)$. ``diyarahman99'' is the third most central user of May network with $(in-degree = 176, out -degree = 1)$ and ``maryamnsharif'' with $(in-degree = 59, out -degree = 0)$ is the fourth most central user of May network.
 
In June, second, fourth and sixth most central users are ``dunyanews'', ``imrankhanpti'' and ``ptiofficial'' with $(in-degree = 290, out -degree = 17), (in-degree = 105, out -degree = 0)$ and $(in-degree = 80, out -degree = 1)$ respectively playing role as \textit{influencers} in the network.
  
In July, ``imrankhanpti'' with $(in-degree = 181, out -degree = 0)$, ``dunyanews'' with $(in-degree = 122, out -degree = 15)$ and ``defencepk'' with $(in-degree = 111, out -degree = 1)$ are the second, fourth and fifth most central user in the network. 

In August, ``Shahzad\_ind'' with $(in-degree = 721, out -degree = 2)$ and ``yadavtejashwi'' with $(in-degree = 622, out-degree = 0)$ are the top two most central users in network.

The overall results of the top three \textit{influencers} of each month is shown in table \ref{tbl:topinfluencers}. 

\begin{figure}[H]
\captionsetup{justification=centering}
\begin{subfigure}{.5\textwidth}
  \centering
  \includegraphics[width=6.0 cm, height= 6.0 cm]{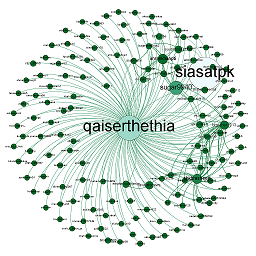}
  \caption{Top ``Influencer'' of March}
  \label{fig:infmarch}
\end{subfigure}%
\begin{subfigure}{.5\textwidth}
  \centering
  \includegraphics[width=6.0 cm, height= 6.0 cm]{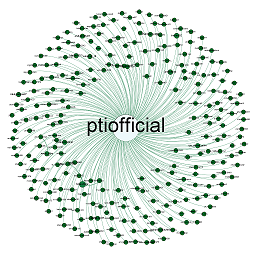}
  \caption{Top ``Influencer'' of April}
  \label{fig:infapril}
\end{subfigure}\hfill
\begin{subfigure}{.5\textwidth}
  \centering
  \includegraphics[width=6.0 cm, height= 6.0 cm]{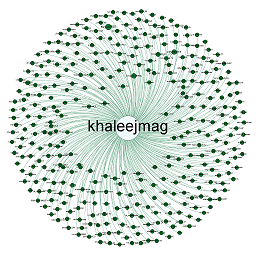}
  \caption{Top ``Influencer'' of May}
  \label{fig:infmay}
\end{subfigure}
\begin{subfigure}{.5\textwidth}
  \centering
  \includegraphics[width=6.0 cm, height= 6.0 cm]{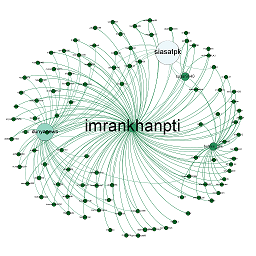}
  \caption{Top ``Influencer'' of June}
  \label{fig:infjune}
\end{subfigure}
\begin{subfigure}{.5\textwidth}
\centering
  \includegraphics[width=6.0 cm, height= 6.0 cm]{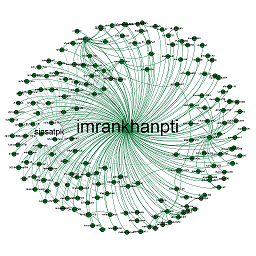}
  \caption{Top ``Influencer'' of July}
  \label{fig:infjuly}
\end{subfigure}
\begin{subfigure}{.5\textwidth}
 \centering
  \includegraphics[width=6.0 cm, height= 6.0 cm]{images/inf/inf-july}
  \caption{Top ``Influencer'' of August}
  \label{fig:infaugust}
\end{subfigure}
\caption{ Connection Pattern for ``Influencer'' of every month. Middle node of every figure is ``Influencer''. Size and color of nodes is according to their degree. Thus, the nodes bigger in size are high degree nodes and smaller nodes are nodes having low degree. Whereas, green  nodes are nodes having lowest degree and color tends to change towards white with increase in degree.}
\label{fig:topinf}

\end{figure}

\begin{table}[H]
\caption{Top Three ``Influencers'' of \#PanamaLeaks network for every month based on ``Betweenness Centrality''}\label{tbl:topinfluencers}
\centering
\begin{tabular}{| p{1.5cm} | p{3cm} | p{2.5cm} | p{2.0cm}| p{3.0cm} | p{1.0cm} |}
\hline
Month &	Name of User &	In-Degree &	Out-Degree & Betweenness	& Rank \\\hline

\multirow{3}{*}{March} 
      & qaiserthethia & 139 &  12 & 315421.518& 3 \\ 
      & alirazatweets & 150 & 1 &251914.020 & 4 \\  
      & xaifoo\_official & 90 & 1 & 200590.403&  5 \\ 
          
\hline

\multirow{3}{*}{April} 
     & ptiofficial & 266 & 1 & 596473.970 &	1 \\ 
   & qaiserthethia & 107 & 12 & 329445.157 & 3 \\  
      & amranafahad	& 116 &	2 &	288166.206 & 4 \\ 
       
\hline

\multirow{3}{*}{May} 
      & khaleejmag & 347 &	1 &	1313558.254	&1  \\ 
      & diyarahman99 & 176 & 1 & 608312.772	& 3\\  
      & maryamnsharif & 59 & 0 & 551766.163 & 4\\ 
        
\hline

\multirow{3}{*}{June} 
      & dunyanews & 290 & 17 & 836335.327 &	2\\ 
      & imrankhanpti & 105 & 0 & 288851.658	& 4 \\  
      & ptiofficial	& 80 & 1 & 213124.288 & 6  \\ 
         
\hline

\multirow{3}{*}{July} 
      & imrankhanpti & 181 & 0 & 800178.225 & 2 \\ 
      & dunyanews & 122 & 15 & 705731.185 & 4 \\  
      & defencepk &	111 & 1 & 601869.039 & 5 \\ 
           
\hline

\multirow{2}{*}{August} 
      & shehzad\_ind & 721 & 2	& 328635.667 & 1 \\ 
      & yadavtejashwi & 622	& 0 & 194985.667 & 2 \\ 
\hline

\end{tabular}
\end{table}
Figure \ref{fig:infmarch} shows the connection pattern of ``qaiserthethia’’, who is playing role as \textit{influencer} in the month of March. Figure \ref{fig:infapril} shows connection pattern ``ptiofficial'', acting as top \textit{influencer} in the month of March. Many arrows are pointing towards this node can be seen in figure \ref{fig:infapril} and most of these nodes are small. Which means mostly isolates like to mention or retweet \textit{influencer }in the network. ``khaleejmag'' is the top \textit{influencer} of May network. Connection pattern of ``khaleejmag'' is shown in \ref{fig:infmay}. June result shows, that \textit{conversation starter} and \textit{active engager} of June both mentioned top \textit{influencer} of that month. However, most nodes connecting with the ``imrankhanpti'' are small, which means they are also isolates having low connectivity in the network. Figure \ref{fig:infjune} \ref{fig:infjuly} 
\ref{fig:infaugust} are showing the connection patterns of top \textit{influencer} of June, July and August respectively.All of the above figures are showing same pattern, which shows connection pattern of \textit{influencer} in the network remain similar.

\begin{figure}[H]
\captionsetup{justification=centering}
\centering
\includegraphics[width=6.0cm, height= 6.0cm]{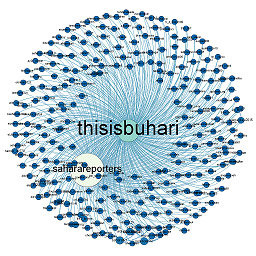}
\caption{Top ``Influencer'' of \#NigeriaDecides case study. Node in the center named as ``thisisbuhari'' is the top  ``Influencer'' of \#NigeriaDecides network. Size and color of nodes is according to their degree. Thus, the nodes bigger in size are high degree nodes and smaller nodes are nodes having low degree. Whereas, blue  nodes are nodes having lowest degree and color tends to change towards white with increase in degree. }\label{fig:infdc}
\end{figure}

In \#NigeriaDecides network, ``thisisbuhari'', ``mbuhari'',``inecnigeria'' are the top three \textit{influencers} ranked by ``betweenness centrality''  playing the role of influencer with $(in-degree = 262, out-degree = 0, betweenness-centrality = 215550.136)$, $(in-degree = 169, out -degree = 0,  betweenness - centrality = 199496.055)$ , $(in-degree = 67, out -degree = 0,  betweenness-centrality = 142580.817)$.
Figure \ref{fig:infdc} shows the top \textit{influencer} ``thisisbuhari'' of          \#NigeriaDecides. Results are consistant with \#PanamaLeaks case study as well as Feng study \cite{feng2016you}. 

\subsection{Active engager}

Result shows that in the network of March second most central user with $(in-degree = 0, out -degree = 65)$ is ``sugar9940'' an \emph{active engager}. In network of April ``sugar9940'' again act as an \textit{active engager} with $(in-degree = 0, out -degree = 41)$ and became fifth most central user in the network. Seventh most central user ``umj3rry'' is \textit{active engager} in network of May with $(in-degree = 0, out -degree = 56)$ and in June it's once again ``sugar9940'' with $(in-degree = 0, out -degree = 96)$. In network of July and August ``lavajatonews'' with $(in-degree = 0, out -degree = 65)$, $(in-degree = 0, out -degree = 14)$ respectively is \textit{active engager}. ``lavajatonews'' is third most central user of July network and seventh most central user of August network playing role as \textit{active engager} (See table \ref{tbl:ae}).

\begin{table}[H]
\captionsetup{justification=centering}
\caption{Presenting ``in-degree", ``out-degree", ``betweeness centrality" and ``Betweeness wise rank"  of the top ``Active Engager'' of each month in the \#PanamaLeaks network.}\label{tbl:ae}
\centering
\begin{tabular}{| l |p{2.3cm}| p{2.0cm} |p{2.0cm} |p{3.0cm}|p{1.7cm}|}
\hline

Month & User name & In-degree & Out-degree & Betweenness & Rank \\
\hline

March &	sugar9940&	0	&65&	356338.123&	2\\
\hline

April	&sugar9940	&0	&41	&216878.111	&5\\
\hline

May&	umj3rry	&0	&56	&357661.301&	7\\
\hline

June	&sugar9940&	0&	96&	530891.138&	3\\
\hline

July &	lavajatonews&	0&	65&	798249.408&	3\\
\hline

August &	lavajatonews&	0&	14	&1360.000&	7\\
\hline

\end{tabular}
\end{table}

\begin{figure}[H]
\captionsetup {justification=centering}
\begin{subfigure}{.5\textwidth}
  \centering
  \includegraphics[width=6.0 cm, height= 6.0 cm]{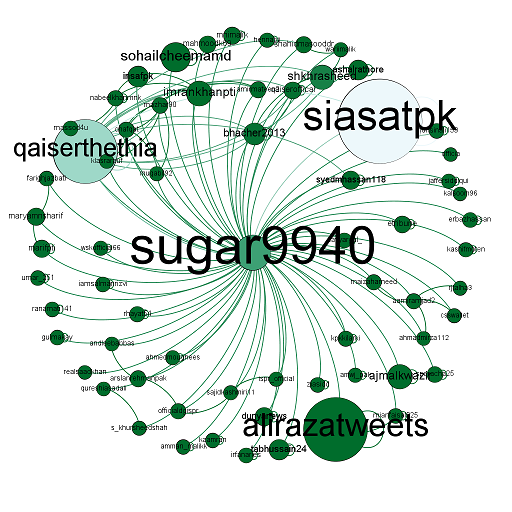}
  \caption{Top ``Active engager'' of March}
  \label{fig:aemarch}
\end{subfigure}%
\begin{subfigure}{.5\textwidth}
  \centering
  \includegraphics[width=6.0 cm, height= 6.0 cm]{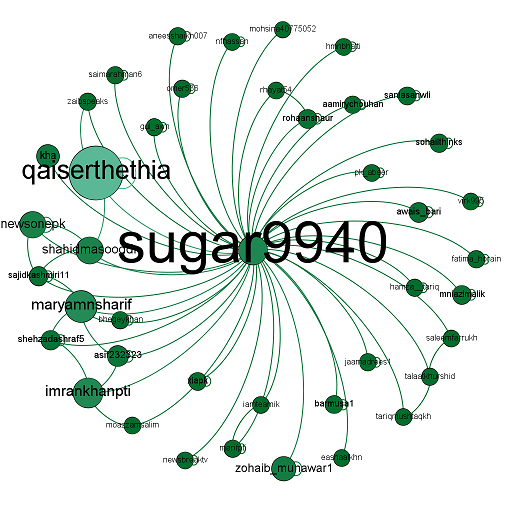}
  \caption{Top ``Active engager'' of April}
  \label{fig:aeapril}
\end{subfigure}\hfill
\begin{subfigure}{.5\textwidth}
  \centering
  \includegraphics[width=6.0 cm, height= 6.0 cm]{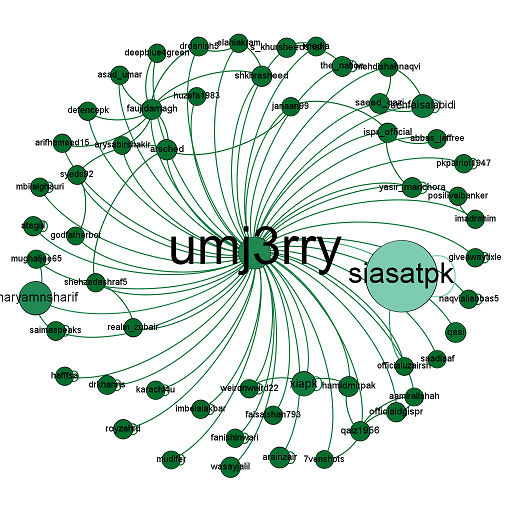}
  \caption{Top ``Active engager'' of May}
  \label{fig:aemay}
\end{subfigure}
\begin{subfigure}{.5\textwidth}
  \centering
  \includegraphics[width=6.0 cm, height= 6.0 cm]{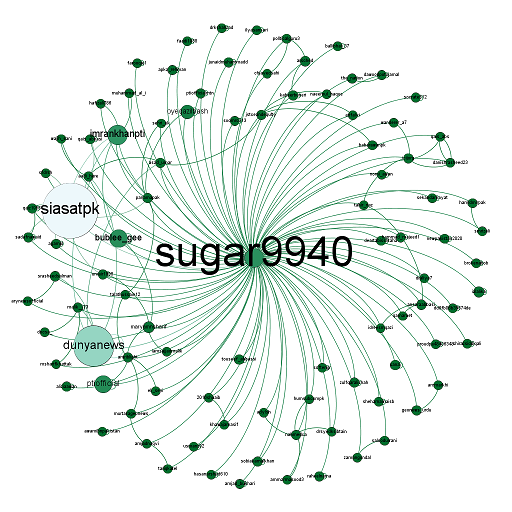}
  \caption{Top ``Active engager'' of June}
  \label{fig:aejune}
\end{subfigure}
\begin{subfigure}{.5\textwidth}
\centering
  \includegraphics[width=6.0 cm, height= 6.0 cm]{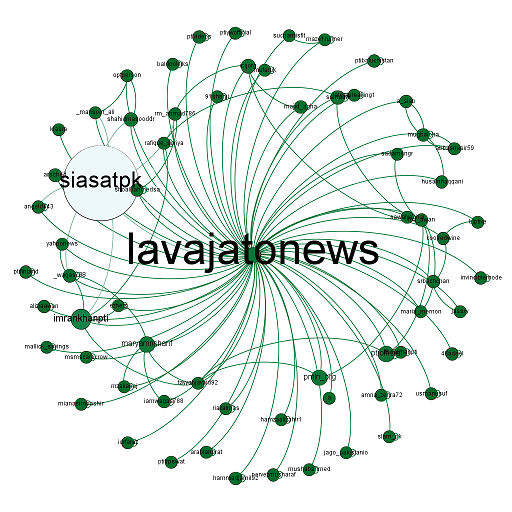}
  \caption{Top ``Active engager'' of July}
  \label{fig:aejuly}
\end{subfigure}
\begin{subfigure}{.5\textwidth}
 \centering
  \includegraphics[width=6.0 cm, height= 6.0 cm]{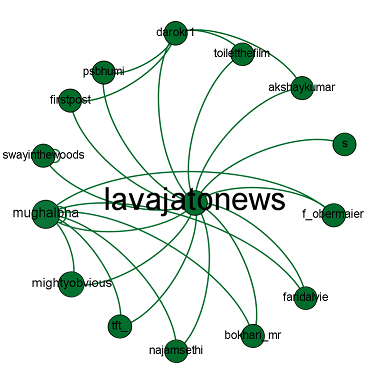}
  \caption{Top ``Active engager'' of August}
  \label{fig:aeaugust}
\end{subfigure}
\caption{Connection Pattern for ``Active engager'' of every month. Middle node of every figure is ``Active engager''. Size and color of nodes is according to their degree. Thus, the nodes bigger in size are high degree nodes and smaller nodes are nodes having low degree. Whereas, green nodes are nodes having lowest degree and color tends to change towards white with increase in degree.}
\label{fig:activeengager}

\end{figure}

Figures \ref{fig:activeengager} is showing connection pattern for \textit{active engagers} from March to August. In March, ``sugar9940'' is playing the role of\textit{ active engager} by retweeting/mentioning other participants of the network. ``sugar9940'' mentiond participant having high degree like ``qaiserthethia'',``alirazatweets'' and ``siasatpk'', connection pattern of March \textit{active engager} can be seen in  \ref{fig:aemarch}.``sugar9940'' is again playing the role of \textit{active engager} in April, Figure  \ref{fig:aeapril} shows the connection pattern of ``sugar9940''. Mentioning ``qaiserthethia'',``maryamnsharif'' and ``imrankhanpti'' three influencers in April network is the reason of his central position in the network. ``umj3rry'' takes the role of \textit{active enager} in May network. Figure \ref{fig:aemay}    shows the connection pattern of ``umj3rry''. By mentioning \textit{conversation starter}, two \textit{influecers} and other participants  in June ``sugar9940'' again is the \textit{active engager} as shown in figure \ref{fig:aejune}. Figure \ref{fig:aejuly} \ref{fig:aeaugust} shows the connection pattern for \textit{active engager} of July and August month respectively. Interesting thing to note is that, in-degree for active engagers of all the months is zero. Which means either no participant has mentioned or retweeted active engager's tweet.       

\begin{figure}[H]
\captionsetup{justification = centering}
\centering
\includegraphics[width=6.0cm, height= 6.0cm]{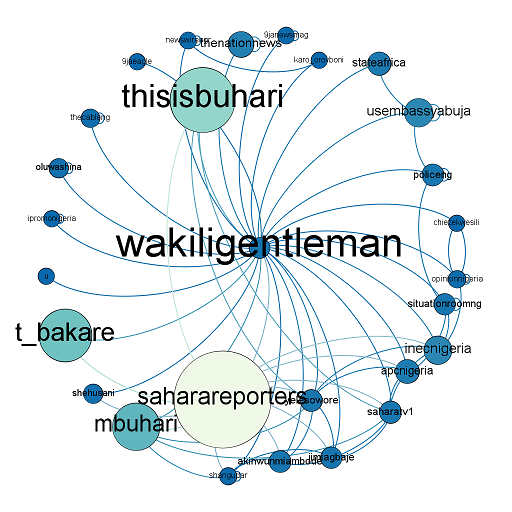}
\caption{Connection Pattern for ``Active Engager'' of \#NigeriaDecides network. Node in the middle of figure is ``Active Engager''. Size and color of nodes is according to their degree. Thus, the nodes bigger in size are high degree nodes and smaller nodes are nodes having low degree. Whereas, blue nodes are nodes having lowest degree and color tends to change towards white with increase in degree.  }\label{fig:aedc}
\end{figure}

``wakiligentleman'' is playing role as \textit{active engager} in the network of \#NigeriaDecides with $in-degree = 1$ and $out-degree = 26$. He is the seventh most central user of the network. ``Betweenness centrality'' of  ``wakiligentleman'' is $112865.285$. By mentioning/retweeting \textit{influencers} as well as \textit{conversation starter} in his tweets, he got high betweenness centrality. Figure \ref{fig:aedc}  shows the connection pattern of ``wakiligentleman'', results are consistant with \#PanamaLeaks case study with many out-going connections. 

\subsection{Network builder}
``shahwar125'' with only $(in-degree = 0, out -degree = 2)$ is interestingly sixth most central user of the March network playing role as \textit{network builder}. With only $(in-degree = 0, out -degree = 7)$ ``ambitiousfree'' is again the sixth most central user of April netowork because he mentioned two influencers ``maryamnsharif'' and ``amranafahad'' in his tweets and build a strategic location in the network like ``shahwar125'' in the network of March month. ``ali\_axhar'' with $(in-degree = 20, out -degree = 8)$, ``deziggnerumer'' with $(in-degree = 0, out -degree = 18)$ are the network builders for month of May and June respectively. In July we could find ``\_waqas788'' as \textit{network builder} on thirteenth position with $(in-degree = 5, out-degree = 21)$. In August with $(in-degree = 0, out -degree = 3)$ ``foroneindia'' is the network builder. He is the fourth most central user in the netwrok. No \textit{influence}r in the network of all months has mentioned \textit{network builders} in his/her tweet but still by connecting two or more Influential,\textit{ network builders} got central positions in the network. The overall results of top \textit{network builder} in the network is shown in table \ref{tbl:nb}. 

\begin{figure}[H]
\captionsetup{justification=centering}
\begin{subfigure}{.5\textwidth}
  \centering
  \includegraphics[width=6.0 cm, height= 6.0 cm]{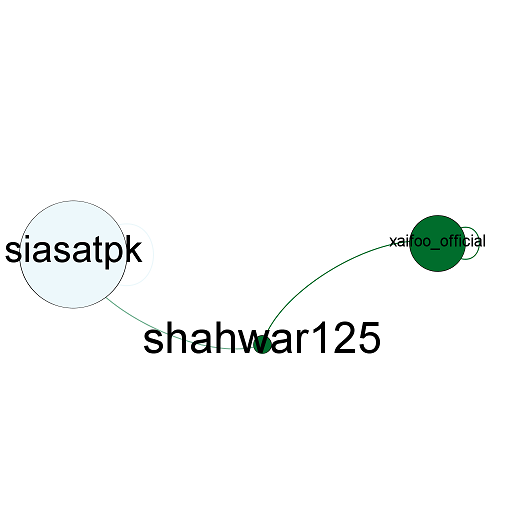}
  \caption{Top ``Network builder'' of March}
  \label{fig:nbmarch}
\end{subfigure}%
\begin{subfigure}{.5\textwidth}
  \centering
  \includegraphics[width=6.0 cm, height= 6.0 cm]{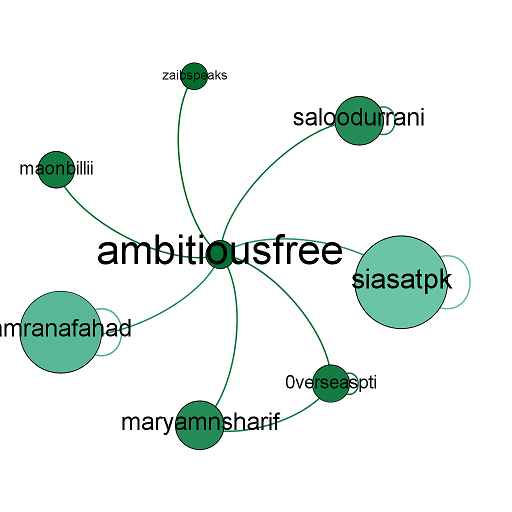}
  \caption{Top ``Network builder'' of April}
  \label{fig:nbapril}
\end{subfigure}\hfill
\begin{subfigure}{.5\textwidth}
  \centering
  \includegraphics[width=6.0 cm, height= 6.0 cm]{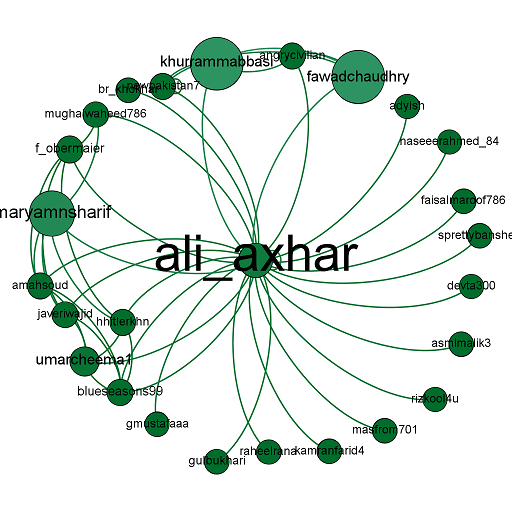}
  \caption{Top ``Network builder'' of May}
  \label{fig:nbmay}
\end{subfigure}
\begin{subfigure}{.5\textwidth}
  \centering
  \includegraphics[width=6.0 cm, height= 6.0 cm]{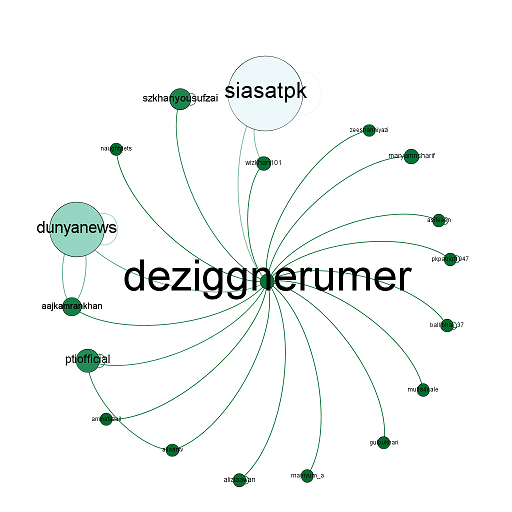}
  \caption{Top ``Network builder'' of June}
  \label{fig:nbjune}
\end{subfigure}
\begin{subfigure}{.5\textwidth}
\centering
  \includegraphics[width=6.0 cm, height= 6.0 cm]{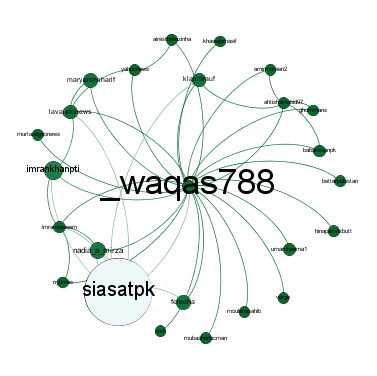}
  \caption{Top ``Network builder'' of July}
  \label{fig:nbjuly}
\end{subfigure}
\begin{subfigure}{.5\textwidth}
 \centering
  \includegraphics[width=6.0 cm, height= 6.0 cm]{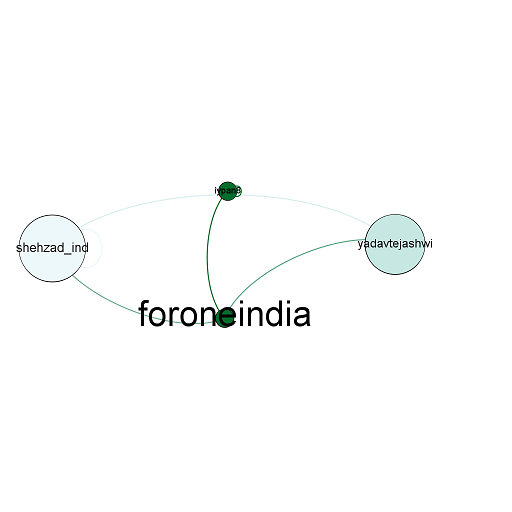}
  \caption{Top ``Network builder'' of August}
  \label{fig:nbaugust}
\end{subfigure}
\caption{Connection Pattern for ``Network Builder'' of every month. Middle node of every figure is ``Network builder''. Size and color of nodes is according to their degree. Thus, the nodes bigger in size are high degree nodes and smaller nodes are nodes having low degree. Whereas, green nodes are nodes having lowest degree and color tends to change towards white with increase in degree.  }
\label{fig:networkbuilders}

\end{figure}

\begin{table}[H]
\captionsetup{justification=centering}
\caption{Presenting ``in-degree'', ``out-degree'', ``betweeness centrality'' and ``rank''  of top ``Network Builder'' of each month in the \#Panamaleaks network based on ``Betweenness Centrality''.}\label{tbl:nb}
\centering
\begin{tabular}{| l |p{2.5cm}| p{2cm} |p{2.0cm} |p{3.0cm}|p{2.0cm}|}
\hline

Month & User name & In-degree & Out-degree & Betweenness & Rank \\
\hline
March	&shahwar125&	0	&2	&186462.694&	6\\
\hline
April	&ambitiousfree&	0	&7	&162629.964	&6\\
\hline
May&	ali\_axhar&	20	&8	&295700.335&	10\\
\hline
June&	deziggnerumer&	0&	18&	267362.714&	5\\
\hline
July &	\_waqas788&	5&	21	&276354.023	&13\\
\hline
August &	foroneindia&	0&	3	&2880.000&	4\\
\hline
\end{tabular}
\end{table}
Node in the middle of Figure \ref{fig:nbmarch} ``shahwar125'' is \textit{network builder} of March. With only two out-degree links, he is the sixth most central user in network. ``shahwar125'' built a strategic location in network by mentioning/retweeting only two participants having high degree. Figures \ref{fig:nbapril} \ref{fig:nbmay} \ref{fig:nbjune} \ref{fig:nbjuly} \ref{fig:nbaugust} shows the connection pattern for \textit{network builder }of April, May, June, July and August respectively. All the above figures show similar pattern of communication. All network builders have few ``in-degree'' and ``out-degree'' links; however, they build strategic location in the \#PanamaLeaks network by mentioning \textit{influencers} or \textit{conversation starter}.

\begin{figure}[H]
\centering
\includegraphics[width=6.0cm, height= 6.0cm]{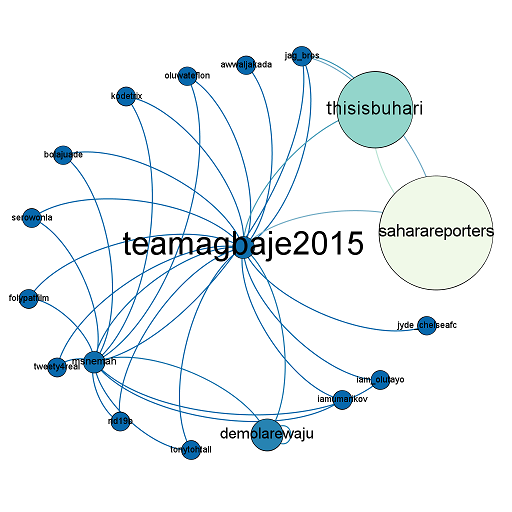}
\caption{Connection Pattern for ``Network Builder'' of \#NigeriaDecides network. Node in the middle of figure is ``Active Engager''. Size and color of nodes is according to their degree. Thus, the nodes bigger in size are high degree nodes and smaller nodes are nodes having low degree. Whereas, blue nodes are nodes having lowest degree and color tends to change towards white with increase in degree.}\label{fig:nbdc}
\end{figure}

``teamagbaje2015'' is playing the role as \textit{network builder} in \#NigeriaDecides network with $in-degree = 13$ and $out-degree = 3$. He is the fifth most centeral user of \#NigeriaDecides network by only mentioning two \textit{influencer} ``thisisbuhari'', ``demolarewaju'' and \textit{conversation starter} ``saharareporters'' of the network. This result is also consistent with \#PanamaLeaks network. Figure \ref{fig:nbdc} is showing the links of network builder of \#NigeriaDecides network.

\subsection{Information bridge} 
Results reveal that there is no\textit{ information bridge} in \#PanamaLeaks six months network. \textit{Information bridge} is the user, who mention/retweets \textit{influencer} tweet and that same tweet is again retweeted by \textit{active engager} \cite{feng2016you}. We could not find any user with few ``in-degree'' and ``out-degree'' links, who has retweeted influencer tweet and then that same tweet is retweeted by \textit{active engager}. We applied this aproach to \#NigeriaDecides case study \cite{udanor2016determining} for validation purpose and according to the results; there exist no user like \textit{information bridge}. Although, we were able to find \textit{conversation starter}, \textit{influencer}, \textit{active engager} and \textit{network builder} in \#NigeriaDecides case study also.

\subsection{Structural analysis} 
\begin{figure}[H]
\captionsetup{justification = centering}
\begin{subfigure}{.5\textwidth}
  \centering
  \includegraphics[width=7.0 cm, height= 6.0 cm]{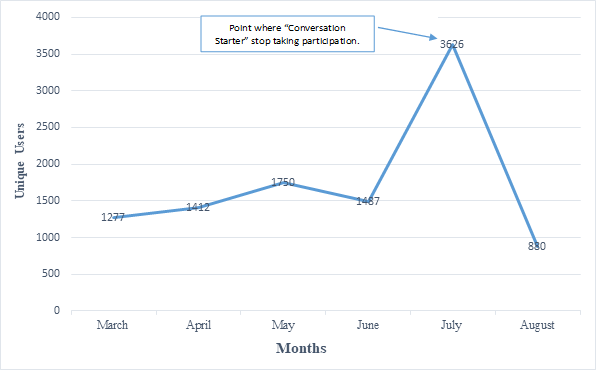}
  \caption{Unique users}
  \label{fig:plot uniqueusers}
\end{subfigure}%
\begin{subfigure}{.5\textwidth}
  \centering
  \includegraphics[width=7.0 cm, height= 6.0 cm]{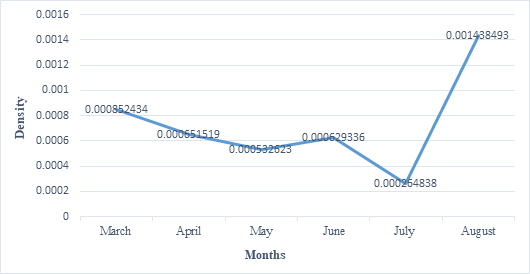}
  \caption{Density}
  \label{fig:plotdensity}
\end{subfigure}\hfill\\
\\
\begin{subfigure}{.5\textwidth}
  \centering
  \includegraphics[width=7.0 cm, height= 6.0 cm]{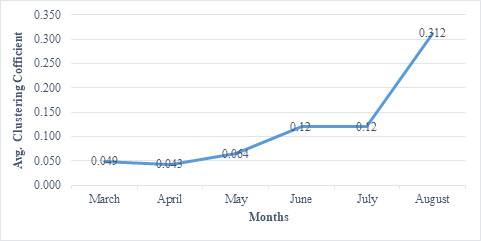}
  \caption{Avg. Clustering coefficient}
  \label{fig:plot clustcoefficient}
\end{subfigure}
\begin{subfigure}{.5\textwidth}
  \centering
  \includegraphics[width=7.0 cm, height= 6.0 cm]{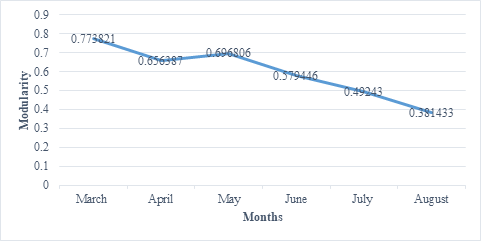}
  \caption{Modularity}
  \label{fig:plotmodularity}
\end{subfigure}

\caption {Number of Unique users are shown in \ref{fig:plot uniqueusers}. Overall Density of \#PanamaLeaks network is shown in \ref{fig:plotdensity}. \ref{fig:plot clustcoefficient} is showing Average Clustering Cofficient of the network. Figure \ref{fig:plotmodularity} is showing Modularity of the netwrok.}
\label{fig:plotsssssss}

\end{figure}

Figure \ref{fig:plot uniqueusers} shows the number of unique users participating in \#PanamaLeaks online discussion every month. Number of unique users coming into \#PanamaLeaks discussion network was $1277$ in March. In April and May $1412$ and $1750$ unique users came into the network respectively. $1487$ unique users joined discussion in June and $3626$ joined discussion in July. However, there is a sudden decrease in number of unique users participating in \#PanamaLeaks discussion in August with $880$ new unique users coming into the network. In September, there are negligible number of users discussing the issue of Panama Leaks anymore. That is the reason we only analyzed this network until August. Figure \ref{fig:plotdensity} shows the density of \#PanamaLeaks network. Gradual decrees in the density of the network can be seen. However, there is a sudden increase in density after July. Average clustering coefficient of \#PanamaLeaks from March to August is shown in Figure \ref{fig:plot clustcoefficient}. Average clustering coefficient of the network starts from $0.049$ on March and with gradual increase ends at $0.312$ on August. Modularity of the \#PanamaLeaks network is $0.773821, 0.656387, 0.696806, 0.579446, 0.49243$ and $0.381433$ for March, April, May, June, July and August respectively (See Figure \ref{fig:plotmodularity}).
\\

\begin{figure}[H]
\begin{subfigure}{.5\textwidth}
  \centering
  \includegraphics[width=6.0 cm, height= 6.0 cm]{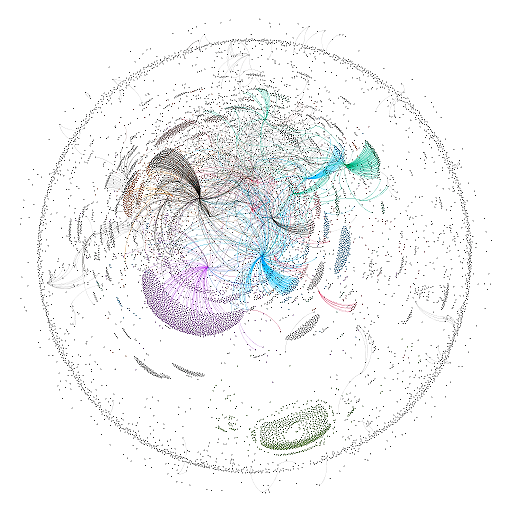}
  \caption{March}
  \label{fig:strmarch}
\end{subfigure}%
\begin{subfigure}{.5\textwidth}
  \centering
  \includegraphics[width=6.0 cm, height= 6.0 cm]{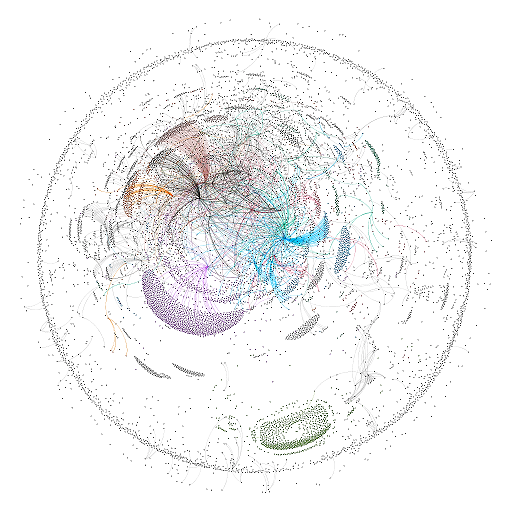}
  \caption{April}
  \label{fig:strapril}
\end{subfigure}\hfill
\begin{subfigure}{.5\textwidth}
  \centering
  \includegraphics[width=6.0 cm, height= 6.0 cm]{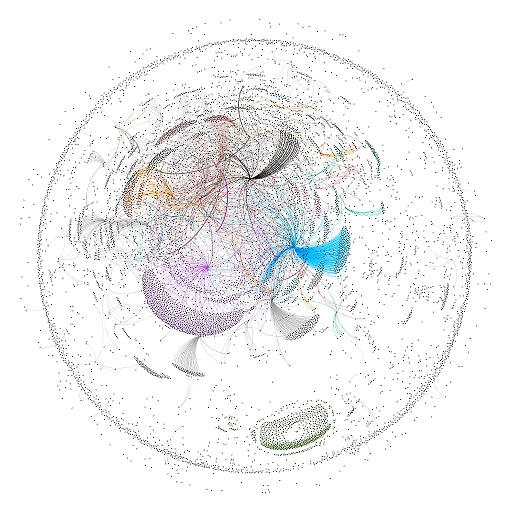}
  \caption{May}
  \label{fig:strmay}
\end{subfigure}
\begin{subfigure}{.5\textwidth}
  \centering
  \includegraphics[width=6.0 cm, height= 6.0 cm]{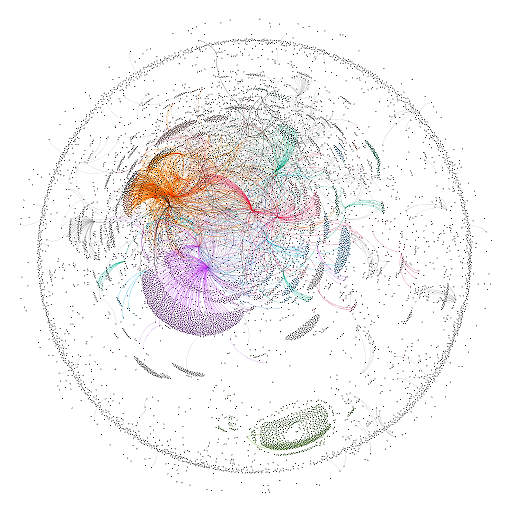}
  \caption{June}
  \label{fig:strjune}
\end{subfigure}
\begin{subfigure}{.5\textwidth}
\centering
  \includegraphics[width=6.0 cm, height= 6.0 cm]{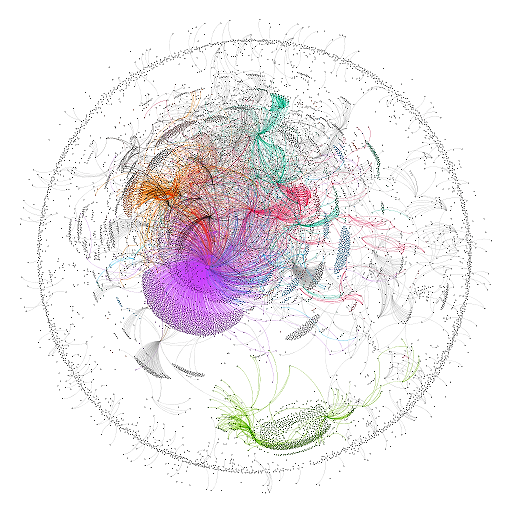}
  \caption{July}
  \label{fig:strjuly}
\end{subfigure}
\begin{subfigure}{.5\textwidth}
 \centering
  \includegraphics[width=6.0 cm, height= 6.0 cm]{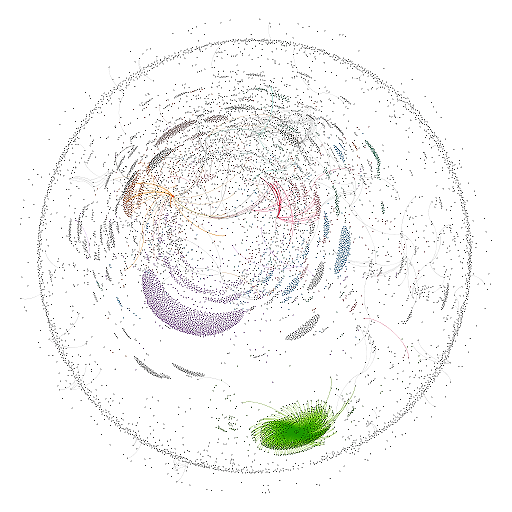}
  \caption{August}
  \label{fig:straugust}
\end{subfigure}

\caption{Dynamic Graph Visualization of \#PanamaLeaks network for each by Gephi using Yifan Hu algorithm. Communities forming around ``Conversation Starter'' and ``Influencers'' are presented in different colors. }\label{fig:monthwisestructure}

\end{figure}

\begin{figure}[H]
\centering
\includegraphics[width=12.31cm, height= 11.25cm]{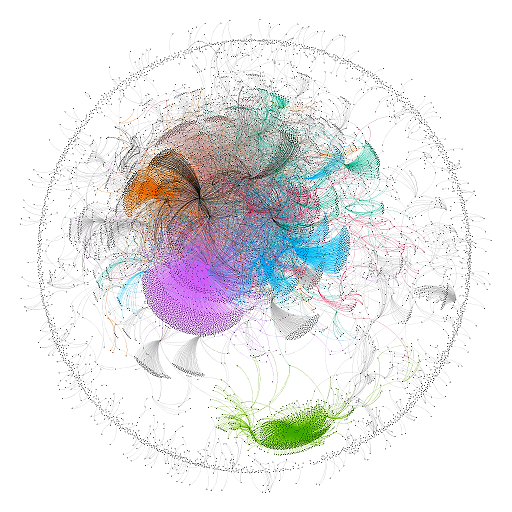}
\caption{Overall network structure of complete six months of  \#PanamaLeaks using Gephi.}\label{fig:overalstructure}
\end{figure}

Figure \ref{fig:monthwisestructure} shows interaction patterns of users, there exist many isolates pointing towards the \textit{influencers} and \textit{conversation starter} in the network. Figure \ref{fig:strmarch} shows the network of March and communities forming around \textit{conversation starter} and \textit{influencers} can be seen. These communities keep on growing until July, which can be seen in figure \ref{fig:strapril}, \ref{fig:strmay}, \ref{fig:strjune}, \ref{fig:strjuly} and \ref{fig:straugust}. However, in August when major influencers and conversation starter left the discussion, communities started to dissolve which can be seen in figure \ref{fig:straugust}. A combine visualization of \#PanamaLeaks network from March to August is shown in figure \ref{fig:overalstructure}. A circle of participants around the graph can be seen. These participants are not connected to any \textit{influencer} or \textit{conversation starter} in the \#PanamaLeaks network.

\section{Discussion}\label{sec:disc}

\subsection{Conversation starter} 
\begin{figure}[H]
\captionsetup{justification=centering}
\centering
\includegraphics[width=10cm, height= 6cm]{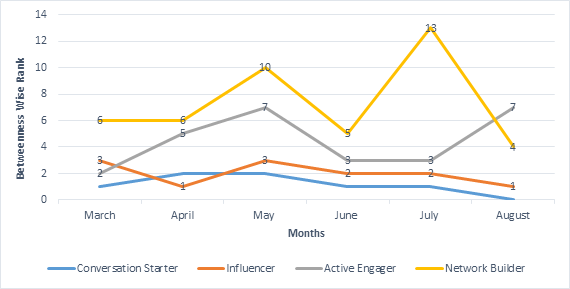}
\caption{Changes in Ranking of Central user’s from March 2017 to August 2017. Months are shown on x-axis and betweenness wise rank are shown along with y-axis.}\label{fig:plots betweenees wise rank}
\end{figure}

A \textit{conversation starter} is one, who is center of the network and connects many ``isolates'' who otherwise have few or no links in the network \cite{feng2016you}. These are the participants with high betweenness \cite{xu2017longitudinal}. Our results reveled the same as many ``isolates'' retweeted or mentioned ``siasakpk'' in their tweets. There were some influential mentioning \textit{conversation starter} in their tweets but they are very few, mostly ``isolates'' tends to connect with \textit{conversation starter} to gain information.     Twitter users with high connectivity are better at influencing information flow in Twitter network \cite{xu2014predicting}. This makes the \textit{conversation starter} center of the network by various incoming connections. However, when \textit{conversation starter} stops taking participation in the network (Stop Tweeting about the Topic) ``isolates'' also lose interest in the online conversation. As results reveled that in August when \textit{conversation starter} is no more in the network, there is sudden decrease in the people discussing about the topic online. However, until the \textit{conversation starter} keeps on participating in the discussion. \textit{Conversation starter} remains on the top central positions in the network.(See Figure \ref{fig:plots betweenees wise rank}).

\subsection{Influencer }
\begin{figure}[H]
\centering
\includegraphics[width=12cm, height= 6cm]{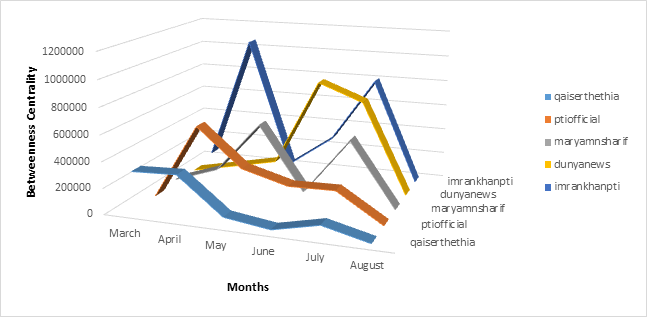}
\caption{Changes in ``Betweenness Centrality'' of selected ``Influencers'' in \#PanamaLeaks network.}\label{fig:plots betweenees}
\end{figure}

There exist \textit{influencers} in the online discussion network \cite{feng2016you} with plentiful ``in-degree'' and few ``out-degree'' links. They are called opinion leaders in the network and isolates like to mention them in their tweets. Isolates also like to retweet opinion leader's tweet and these opinion leaders are usually mass media organizations or celebrities in different fields of life \cite{feng2016you}. Opinion leaders create many tweets to influence other users in online discussion. In this way, they act to influence opinion in an online discussion network. Information is going from  opinion leaders to users with less connectivity and this supports the argument by \cite{dahlberg2007rethinking,murdock2004dismantling} that opinion leaders by using their network connectivity replicate asymmetric offline power in online discussion. The participant like influencer who has high degree in network act as hub in the network \cite{ahuja2003individual}. However, Opinion leader's influence is not consistent as the results show in \#PanamaLeaks study their influence can increase/decrease with the passage of time. There can be changes in the``betweenness centrality'' of network influencers as well. Figure \ref{fig:plots betweenees}
 shows the changing in betweenness centrality of five selected \textit{influencers} in the \#PanamaLeaks network.

\subsection{Active engager }
A user with many ``out-degree'' and none or few ``in-degree'' links is \textit{active engager} in the network \cite{feng2016you}. Results also reveal that there always exist an \textit{active engager }in the network from start of conversation until end. However, position of \textit{active engager} is not consistent and other \textit{active engager} can take his/her place with the passage of time. \textit{Active engager} is also among the top ten most central users of the network. Results show that \textit{active engager} is one who is active and with intention to get information. However, there are usually few or no user in the network to follow or mention him in their tweets. From the centrality point of view,\textit{ active engager} is an important user. However, \textit{Active engager} in a socity is a person with special knowledge, who has potential to become influencer \cite{quercia2011our}. 

\subsection{Network builder} 
\textit{Network builder} in a network has few or none ``in-degree'' and few ``out-degree'' links \cite{feng2016you}. His basic role in the network is to associate two or more influential nodes in network. By this, \textit{network builder} builds a strategic location in network due to which it gets high centrality.  However, in \cite{feng2016you} ``wnkr'' is a \textit{network builder} and its in-degree is ``0''. Which means no one either has mentioned or retweet ``wnkr'' in their tweets. In \#PanamaLeaks  network, \textit{network builders} follow the same tendency. In this context, we can conclude that \textit{network builder} do help in creating connection between two or more \textit{influencers} in the online discussion network. However, \textit{influencers} usually do not mention or reply on their tweets. Therefore, their impact on the opinion in a network is petite.

\subsection{Information bridge }
\textit{Information bridge} is a user with few ''in-degree'' and ``out-degree'' links, who connects an \textit{influencer} and \textit{active engager} as noted in \cite{feng2016you}. However, results of six months of network shows that there is no user such as \textit{information bridge} exist in the network. We  implemented this concept on the dataset \#NigeriaDecides used in \cite{udanor2016determining} for validation purpose. Result shows that in the \#NigeriaDecides, there does not exist any user who has retweeted \textit{influencer} tweet and that tweet is retweeted by \textit{active engager}. Active engager's basic role in network is to get engage with other users \cite{feng2016you}, and for this purpose; \textit{active engager} likes to retweet mostly \textit{influencer/conversation starter} tweets. Active engagers do not mostly retweet/mention ``isolates'' in their tweets. In this context, we can conclude that there is very less chance that an \textit{active engager }retweets same tweet in which a person with few ``in-degree'' and ``out-degree'' has mentioned \textit{influencer}.

\subsection{Network structure} 
Unique users have increased until June and then there is a sudden decrease in the number of unique users as shown in figure \ref{fig:plot uniqueusers}. When \textit{conversation starter} stop taking participation in the online discussion there is a sudden decrease in number of unique users participation. Users with high connectivity like \textit{conversation starter} can influence the participation of users \cite{zhao2017factors}. In this context, opinion leaders have the power to keep isolates interested in the online discussion. Network density can decrease with the number of increase users in the network and increase when number of participant’s decrease \cite{lee2017mapping}, density of the network is showing same tendency. Participant of the online discussion tends to cluster around the influential participants and with the passage of time in presence of \textit{conversation starter} and \textit{influencers}, participant became less independent as modularity decreased \cite{xu2017longitudinal}. However, with many participants having low degree pointing towards the \textit{conversation starter} and \textit{influencer}. We can conclude that they play role as hub in the online discussion network.

\newpage
\textbf{Reference}
\bibliography{sample}
\bibliographystyle{unsrt}

\end{document}